\documentclass[runningheads]{llncs}
\usepackage{graphicx}
\usepackage{amssymb,amsmath,amsfonts}

\usepackage[english]{babel}
\newcommand {\twofl}[2]{\stackrel {\mbox {\scriptsize #1}}{#2}}
\newcommand{\algend}{\rule{1em}{0pt}$\blacksquare$}
\newcommand{\trans}{\textbf{\emph{Trans}}}

\newcommand{\RG}{\overset{\rightarrow}{R}F}

\begin{document}
\title{A polynomial-time algorithm for the routing flow shop problem with two machines: an asymmetric network with a fixed number of nodes\thanks{This research was supported by the program of fundamental scientific researches of
the SB RAS No I.5.1., project No 0314-2019-0014, and by the Russian Foundation for
Basic Research, projects 20-07-00458 and 18-01-00747.}}
\titlerunning{On the asymmetric two-machine routing flow shop}
%
\author{Ilya Chernykh\inst{1,2,3}\orcidID{0000-0001-5671-8562} \and
Alexander Kononov\inst{1,2}\orcidID{0000-0001-6144-0251} \and
Sergey Sevastyanov\inst{1}\orcidID{0000-0003-0347-1396}}
\authorrunning{I. Chernykh, A. Kononov and S. Sevastyanov}
%
\institute{Sobolev Institute of Mathematics, Koptyug ave. 4, Novosibirsk, 630090, Russia
\email{\{idchern,alvenko,seva\}@math.nsc.ru}
\and
Novosibirsk State University, Pirogova str. 2, Novosibirsk, 630090, Russia
\and
Novosibirsk State Technical University, Marksa ave. 20, Novosibirsk, 630073, Russia
}
\maketitle              
\begin{abstract}
We consider the routing flow shop problem with two machines on an asymmetric network. For this problem we discuss properties of an optimal schedule and present a polynomial time algorithm assuming the number of nodes of the network to be bounded by a constant. 
To the best of our knowledge, this is the first positive result on the complexity of the routing flow shop problem with an arbitrary structure of the transportation network, even in the case of a symmetric network.
This result stands in contrast with the complexity of the two-machine routing open shop problem, 
which was shown to be NP-hard even on the two-node network.

\end{abstract}

\section{Introduction}\label{int}
A flow shop problem to minimize the makespan (also known as {\em Johnson's problem}) is probably the first machine scheduling problem described in the literature \cite{J53}. It can be set as follows.

\medskip\noindent\textbf{Flow shop problem.}\ 
Sets ${\cal M}$ of machines and ${\cal J}$ of jobs are given, each machine $M_i\in{\cal M}$ has to process each job $J_j\in{\cal J}$; such an operation takes $p_{ji}$ time units. Each job has to be processed by machines in the same order: first by machine $M_1$, then $M_2$ and so on. No machine can process two jobs simultaneously. The goal is to construct a feasible schedule of processing all the jobs within the minimum {\em makespan} (which means, with the minimum completion time of the last operation). According to the traditional three-field notation of scheduling problems (see \cite{LaLeRiSh93}), Johnson's problem with a fixed number $m$ of machines is denoted as $Fm||C_{\max}$. 

Problem $F2||C_{\max}$ can be solved to the optimum by the well-known Johnson's algorithm, which basically is a sorting of the set of jobs according to Johnson's rule \cite{J53}. On the other hand, problem $F3||C_{\max}$ is NP-hard in the strong sense~\cite{GaJoSe76}.

In classical scheduling problems (including flow shop), it is assumed that the location of each
machine is fixed, and either there is no pre-specified delay between the
processing of two consecutive operations of a job or such a delay depends on the distance
between the corresponding machines. However, this assumption often diverges from
real-life situations. Imagine that the company is engaged in the
construction or maintenance of country houses, cottages or chalets. The
company has several crews which, for example, specialize either in
preparing the site for construction, or filling the foundation, or
building a house, or landscaping the site. The facilities are located in
a suburban area, and each team must move from place to place to carry
out their work. The sequence of jobs performed by various crews is
fixed, e.g., you cannot start to build a house before filling the foundation.

To take into account the situation described above, we consider a natural combination of $Fm||C_{\max}$ with the well-known traveling salesman problem, a so-called {\em routing flow shop problem} introduced in \cite{AvBe96}. In this model, jobs are located at nodes of a transportation network $G$, while machines have to travel over the edges of the network to visit each job and perform their operation in the flowshop environment. All machines start from the same location (the {\em depot}) and have to return to the depot after performing all the operations. The completion time of the last machine \emph{action} (either traveling or processing an operation of some job in the depot) is considered to be the \emph{makespan} of the schedule ($C_{\max}$) and has to be minimized. (See Sect. \ref{se1} for the detailed formulation of the problem.) 

We denote the $m$-machine routing flow shop problem as $RFm||C_{\max}$ or $RFm|G=W|C_{\max}$, when we want to specify a certain structure $W$ of the transportation network.

The routing-scheduling problems can simulate many problems in real-world applications. Examples
of applications where machines have to travel between jobs include
situations where parts are too big or heavy to be moved between machines
(e.g., engine casings of ships), or scheduling of robots that perform
daily maintenance operations on immovable machines located in different
places of a workshop \cite{AvBe99}. Another interesting
application is related to the routing and scheduling of museum visitors
traveling as homogeneous groups \cite{YuLinChou10}. The model is embedded in
a prototype wireless context-aware museum tour guide system developed
for the National Palace Museum of Taiwan, one of the top five museums in the world.

The routing flow shop problem is still understudied. Averbakh and Berman \cite{AvBe96} considered $RF2||C_{\max}$ with exactly one job at each node, under the following restriction: each machine has to follow some \textbf{shortest route} through the set of nodes of the network (not necessarily the same for both machines). This will be referred to as an {\em AB-restriction}. They proved that for the two-machine problem the AB-restriction affects the optimal makespan by a factor of at most $\frac32$, and this bound is tight. They also showed that, under this restriction, there always exists a {\em permutation} optimal schedule, in which machines process jobs in the same order (a {\em permutation property}). Using this property, they presented $O(n\log n)$ algorithms for solving $RF2|AB$-restriction$,G=W|C_{\max}$ to the optimum, where $W$ is a tree or a cactus, $n$ is the number of jobs. These algorithms, therefore, provide a $\frac32$-approximation for the problem without the AB-restriction on a tree or on a cactus with a single job at each node. Later on (\cite{AvBe99}), they extended these results to the case of an arbitrary graph $G$ and an arbitrary number of machines $m$ by presenting a $\frac{m+1}2$-approximation algorithm for the $RFm||C_{\max}$ problem. Yu and Znang \cite{YuZh11} improved on the latter result and presented an $O(m^{\frac 23})$-approximation algorithm based on a reduction of the original problem to the permutation flow shop problem. 

A generalized routing flow shop problem with buffers and release dates of jobs was also considered in \cite{JoMa14}. The authors present a heuristic based on solving the corresponding multiple TSP.

Yu {\em et al.} \cite{YuZhWaFa11} investigated the $RF2||C_{\max}$ problem with a single job at each node farther. They obtained the following results:
\begin{enumerate}
	\item The permutation property also holds for the problem without the AB-restriction.
	\item The problem is ordinary NP-hard, even if $G$ is a tree (moreover, if $G$ is a spider of diameter 4 with the depot in the center).
	\item There is a $\frac{10}7$-approximation algorithm that solves the $RF2|G=tree|C_{\max}$ problem in $O(n)$ time.
\end{enumerate}

Finally, the possibility of designing a polynomial-time algorithm for the special case of our problem, when the transportation network is symmetric, was claimed in \cite{CheKoSe18} (although, without any proof). 

In the present paper, we investigate the generalization of $RF2||C_{\max}$ problem to the case of  asymmetric travel times and of an arbitrary number of jobs at any node. Thus, we have to consider a \textbf{directed network} $G$ in which the travel times through an edge may be different in the opposite directions. (We will denote such a problem by $\RG 2||C_{\max}$.) We prove that the permutation property holds for this version of the problem, as well. We also establish another important property: there exists an optimal permutation schedule (with the same job processing order $\pi$ on both machines) such that for each node $v$, sub-sequence $\pi_v$ of $\pi$ consisting of all jobs from node $v$ obeys Johnson's rule. These two properties allow us to design a dynamic programming algorithm which solves this problem in time $O(n^{g^2+1})$, where $g$ is the number of nodes in $G$. Thereby, we have established a polynomial-time solvability of the asymmetric two-machine routing flow shop problem with a constant number of network nodes. This result stands in contrast with the complexity result for the two-machine routing open shop problem, which is known to be ordinary NP-hard even if $G$ consists of only two nodes (including the depot) \cite{AvBeCh06}.

The structure of the paper is as follows. Section \ref{se1} contains a formal description of the problem under investigation, as well as some notation and definitions. Properties of an optimal schedule are established at the beginning of Section \ref{se3} which also contains a description of the exact algorithm for solving the problem. The analysis of its qualities follows in Section \ref{se4}. Section \ref{concl} concludes the paper with some open questions for further investigation.

\section{Problem setting, definitions and notation}\label{se1}

Farther, throughout the paper, an expression of the form $x\in[\alpha,\beta]$ (where $\alpha$ and $\beta$ are integers, and $x$ is an integer variable, by definition) means that $x$ takes any integral values from this interval; $[\beta]\doteq\{1,2,\dots,\beta\}$. In this paper we will consider the following problem.

\medskip\textbf{Problem $\RG 2||C_{\max}$.}\ We are given $n$ jobs $\{J_1,\dots,J_n\}$ that are to be processed by two dedicated machines denoted as $A$ and $B$. For each $j\in[n]$, job $J_j$ consists of two operations that should be performed in the given order: first the operation on machine $A$, and then on machine $B$. Processing times of the operations are equal to $a_j$ and $b_j$, respectively. All jobs are located at nodes of a transportation network; the machines move between those nodes along the arcs of that network. At the beginning of the process, both machines are located at a node called a \emph{depot}, and they must return to that very node after completing all the jobs. 

Without loss of generality of the problem (and for the sake of convenience of the further description and analysis of the algorithm presented in Section \ref{se3}), we will assume that a \emph{reduced network} $G=(V,E)\ (|V|=g+2)$ is given, in which: (1) only \emph{active nodes} are retained, i.e., the nodes containing jobs (they will be referred to as \emph{job nodes}) and two \emph{node-depots}: the \emph{start-depot} and the \emph{finish-depot}; (2) there are no jobs in both depots (otherwise, we split the original depot into three copies, the distances between which are equal to zero; one of those copies is treated as a job node, while the other two are job-free); the start-depot and the finish-depot get indices 0 and $g+1$, respectively, while all job nodes get indices $i\in [g]$ ($g$ is the number of job nodes); thus, starting from the start-depot, each machine will travel among the job nodes, and only after completing all the jobs it may arrive at the finish-depot; (3) $G$ is a \textbf{complete directed graph} in which each arc $e=(v_i,v_j)\in E$ is assigned a non-negative weight $\rho(e)=\rho_{i,j}$ representing the shortest distance between the nodes corresponding to $i$ and $j$ in the source network \textbf{in the given direction}; therefore, the weights of arcs satisfy the triangle inequalities; at that, the \textbf{symmetry of the weights is not assumed}, i.e., the weights of the forward and the backward arcs may not coincide. The objective function $C(S)$ is the time, when machine $B$ arrives at the finish-depot in schedule $S$, and this time should be minimized.

\medskip Other designations: $\mathbf{N}\doteq(n_1,\dots,n_g)$, where $n_i$ denotes the number of jobs located at job node $i\in[g]$.\ $\|K\|_1\doteq \sum_{i\in[g]} |k_i|$ denotes the \emph{1-norm} of vector $K=(k_1,\dots,k_g)$.

Given an integer $d>0$, we define a partial order $\lessdot$ on the set $R^d$ of $d$-dimensional real-valued vectors, such that for any two vectors $x'=(x_1',\dots,x_d')$, $x''=(x_1'',\dots,x_d'')\in R^d$ the relation $x'\lessdot x''$ holds, if and only if $x_i'\le x_i'',\ \forall\ i\in[d]$. By ${\cal J}(v)$, we will denote the set of indices of jobs located at node $v\in V$. 

By a \emph{schedule}, we will mean, as usual, the set of starting and the completion times of all operations. Since, however, such a schedule model admits a continuum set of admissible values of its parameters, it will be more convenient for us to switch to a discrete model in which any schedule is determined by a pair of permutations $\{\pi',\pi''\}$ specifying the orders of processing the jobs by machines $A$ and $B$, respectively. Each pair $(\pi',\pi'')$ uniquely defines both the routes of the machines through the nodes of network~$G$ and an \emph{active schedule} $S(\pi',\pi'')$ of job processing which is defined as follows.

A schedule $S(\pi',\pi'')$ is called \emph{active}, \emph{iff}: (1) it is feasible for the given instance of problem $\RG 2||C_{\max}$; (2) it meets the precedence constraints imposed by permutations $\{\pi',\pi''\}$; (3)~the starting time of no operation in this schedule can be decreased without violating the above mentioned requirements. 

An active schedule $S(\pi',\pi'')$ is called a \emph{permutation} one, if $\pi'=\pi''$.

\begin{definition}\label{de1}\rm
For each $j\in[n]$, we define a \emph{priority vector} $\chi_j=(\chi_j',\chi_j'',j)$ of job $J_j$, where $(\chi_j'=1,\ \chi_j''=a_j)$, if $a_j\le b_j$, and $(\chi_j'=2,\ \chi_j''=-b_j)$, otherwise. We next define a strict linear order $\prec$ on the set of jobs: for two jobs $J_j,J_k\ (j,k\in[n])$ the relation $J_j\prec J_k$ holds, \emph{iff}\ $\chi_{j}<_{\mbox{\rm\scriptsize lex}}\chi_{k}$ (i.e., vector $\chi_{j}$ is lexicographically less than $\chi_{k}$). Clearly, for any two jobs $J_j,J_k\ (j\ne k)$, one and only one of two relations holds: either $J_j\prec J_k$ or $J_k\prec J_j$.

We will say that a permutation of jobs $\pi$ and the corresponding permutation schedule meet the \emph{Johnson local property}, if for each node $v\in V$ the jobs from ${\cal J}(v)$ are sequenced in permutation $\pi$ \emph{properly}, which means: in the lexicographically increasing order of their priority vectors. (Johnson~\cite{J53} showed that in the case of the networkless two-machine flow shop problem, such a job order $\pi$ provides the optimality of the corresponding permutation schedule.)
\end{definition}

\section{Properties of the optimal schedule and an algorithm for the exact solution of problem $\RG 2||C_{\max}$}
\label{se3}

The algorithm described in this section is based on two important properties of the optimal schedule established in the following theorems.

\begin{theorem}\label{th1}
For any instance $I$ of problem $\RG 2||C_{\max}$ there exists an optimal schedule which is a permutation one.
\end{theorem}

\begin{theorem}\label{th2}
For any instance $I$ of problem $\RG 2||C_{\max}$ there exists a permutation schedule which meets the Johnson local property and provides the minimum makespan on the set of all permutation schedules.
\end{theorem}

The proofs of these theorems are omitted due to the volume limitations. They can be found in Appendix.
Two theorems above imply the following 

\begin{corollary}\label{co1}
For any instance $I$ of problem $\RG 2||C_{\max}$ there exists an optimal schedule which is a permutation one and meets the Johnson local property.
\end{corollary}

The algorithm for computing the exact solution of problem $\RG 2||C_{\max}$ is based on the idea of Dynamic Programming and on the two properties of optimal solutions mentioned in Corollary \ref{co1} (and, thus, enabling us to restrict the set of schedules under consideration by job sequences which meet these properties). So, from now on, we will consider only permutation schedules which meet the Johnson local property.

Let us number the jobs at each node $v_i$ \emph{properly}, i.e., in the ascending order of the relation $\prec$ (see Definition \ref{de1}, p. \pageref{de1}). Then, due to Theorem \ref{th2}, jobs at each node $v_i\ (i\in[g])$ should be processed in the order $\pi_i=(1,2,\dots,n_i)$. According to this order, the jobs at node $v_i$ will be numbered by two indices: $J_{ij}\ (j\in[n_i])$.

In the schedule under construction, we will highlight the time moments when a machine $M\in\{A,B\}$ completes a portion of jobs at node $v_i$ and is preparing to move to another node. Each such moment will be called an \emph{intermediate finish point} of machine $M$ or, in short, an \emph{if-point} of machine $M$. It follows from Theorem \ref{th2} that at each if-point $t'$ of machine $A$ the set of jobs already completed by the machine is a collection of some \textbf{initial segments} $[1,\dots,k_i]$ of sequences $\{\pi_i\,|\,i\in[g]\}$. This collection can be specified by a $g$-dimensional integral vector $K=(k_1,\dots,k_g)$ (and will be denoted as ${\cal J}(K)$), where $k_i$ denotes the number of jobs performed by machine $A$ at node $i$ by time $t'$. 

By Theorem~\ref{th1}, machine $B$ completely reproduces the route of machine $A$ through network nodes (as well as the order of processing the jobs by that machine) and, at some (later) point in time $t''\ge t'$, it also finds itself at its \emph{if}-point with \textbf{the same set} ${\cal J}(K)$ of completed jobs, defined by vector $K$. Thus, a natural correspondence is established between the \emph{if} -points of machines $A$ and $B$: they are combined into pairs $({t_s}',{t_s}'')$ of \emph{if}-points at which the sets of jobs completed by machines $A$ and $B$ coincide and are defined by the same vector $K_s=(k_1^s,\dots,k_g^s)$. The pairs of \emph{if}-points divide the whole process of performing the jobs by machines $A$ and $B$ into \emph{steps} ($s=1,2,\dots,\bar s$), each step $s$ being defined by two parameters: the node index ($i_s$) and the number of jobs ($d_s$) performed in this step at node $i_s$.

The tuple $\widehat{K}\doteq (K,i^*)$ consisting of a value of vector $K=(k_1,\dots,k_g)$ and a value of a node index~$i^*$ determines a \emph{configuration} of a partial schedule of processing the subset of jobs ${\cal J}(K)$, with the final job at node $i^*$. The set of \emph{admissible configurations} is defined as the set including all \emph{basic configurations} (with values $K=(k_1,\dots,k_g)\in [0,n_1]\times\dots\times[0,n_g],\ i^*\in[g]$, such that $k_{i^*}>0$), as well as two \emph{special configurations}: the \emph{initial} one $\widehat{K}_S=(\mathbf{0},0)$ and the \emph{final} one $\widehat{K}_F= (\mathbf{N},0)$.

Algorithm ${\cal A}_{DP}$ for constructing the optimal schedule makes two things: 1)~it enumerates all possible configurations of partial schedules, and 2) for each of them, it accumulates the maximum possible set of pairwise incomparable solutions (characterized by pairwise incomparable pairs $(t',t'')$ of \emph{if}-points with respect to the relation~$\lessdot$). In other words, given a configuration $\widehat{K}\doteq (K,i^*)$, we consider a ``partial'' bi-criteria problem ${\cal P}(\widehat{K})$ of processing the jobs from ${\cal J}(K)$, with the final job at node $i^*$. The objective is to minimize the two-dimensional vector-function $\bar F\doteq (F_1,F_2)$, where $F_1,F_2$ are the completion times of jobs from ${\cal J}(K)$ by machines $A$ and $B$, respectively. We compute the complete set ${\cal F}(\widehat{K})$ of representatives of Pareto-optimal solutions of this problem.

For each solution $\bar F=(F_1,F_2)\in {\cal F}(\widehat{K})$, let us define the parameter $\Delta(\bar F)=F_2-F_1$. The set ${\cal F}(\widehat{K})$ for each configuration $\widehat{K}$ will be stored as the list sorted in the ascending order of component $F_1$. (At that, the values of $F_2$ and $\Delta(\bar F)$ strictly decrease.) The first element of each list ${\cal F}(\widehat{K})$ will be a solution with the value $F_1=0$. This is either a \emph{dummy solution} $\tilde F=(0,\infty)$ (added to each list ${\cal F}(\widehat{K})$ at the beginning of its formation), or a real solution with the value $F_1=0$ (if it is found).

In the course of the algorithm, configurations $\{\widehat{K}=(K,i^*)\}$ are enumerated (in order to create lists ${\cal F}(\widehat{K})$) in non-decreasing order of the norm $\|K\|_1$ of vectors $K$. At that, the whole algorithm is divided into three stages: the \emph{initial}, the \emph{main} and the \emph{final} one. Configurations with $i^*=0$ are considered in the initial and the final stages only.

In the \textbf{initial stage}, list ${\cal F}(\widehat{K}_S)$ for the \emph{initial configuration} $\widehat{K}_S=(\mathbf{0},0)$ is created. It consists of the single solution $(0,0)$.

In the \textbf{main stage}, for each $k\doteq\|K\|_1=1,\dots,n$, vectors $K$ are enumerated in lexicographical ascending order; for each given vector $K=(k_1,\dots,k_g)$, those values of $i^*\in[g]$ are enumerated only for which $k_{i^*}>0$ holds.

In the \textbf{final stage}, for the \emph{final configuration} $\widehat{K}_F=(\mathbf{N},0)$, we find its optimal solution by comparing $g$ variants of solutions obtained from the optimal solutions of configurations $\{(\mathbf{N},i)\,|\,i\in[g]\}$. For each configuration $\widehat{K}_i=(\mathbf{N},i)$, its optimal solution $\bar F^*_i=(F_1^*,F_2^*)$ (with the minimum value of the component $F_2$) is located at the very end of list ${\cal F}(\widehat{K}_i)$. Having added to $F_2^*$ the distance $\rho_{i,0}$ from node $v_i$ to the depot, we obtain the value of the objective function $C(S)$ of our problem for the given variant of schedule $S$. Having chosen (from $g$ variants) the variant with the minimum value of the objective function, we find the optimum.

To create list ${\cal F}(\widehat{K})$ for a given configuration $\widehat{K}=(K,i^*)$ of the \textbf{main stage}, we enumerate such values of the configuration $\widehat{K}'=(K',i')$ obtained at the completion of the previous step of the algorithm (we will call that configuration a \emph{pre-configuration}, or ``p-c'', for short), that $i'\ne i^*$, and that the vectors $K$ and $K'$ differ in exactly one ($i^*$th) component, so as $k_{i^*}'<k_{i^*}$. At that, if $K'=\mathbf{0}$, then $i'=0$, which means that $\widehat{K}'$ is the \emph{initial configuration}. If, alternatively, $K'\ne\mathbf{0}$, then $k'_{i'}>0$. (Clearly, there is no need for a machine to come to node $v_{i'}$ without doing any job at it.)

We note that for each configuration $\widehat{K}=(K,i^*)$ of the main stage, each variant of its p-c $\widehat{K}'=(K',i')$ can be uniquely defined by the pair $D=(d,i')$, where  $i'\in[0,g]\setminus\{i^*\}$, and $d\in[k_{i^*}]$ is the number of jobs being processed in this step at node $v_{i^*}$. The pairs $(d,i')$ are enumerated so as the loop on $d$ is an exterior one with respect to the loop on $i'$.

For each given value of $d$, we construct an \emph{optimal} schedule $S_d= S(\widehat{K},d)$ in problem $F2||C_{\max}$ for the jobs from ${\cal J}(\widehat{K},d)= \{J_{i^*,j}\,|\,j\in [k_{i^*}-d+1,k_{i^*}]\}$, and then compute three characteristics of that schedule: $L_1(\widehat{K},d)$ and $L_2(\widehat{K},d)$, which are the total workloads of machines $A$ and $B$ on the set of jobs ${\cal J}(\widehat{K},d)$, and also $\delta(\widehat{K},d)=C_{\max}^*(\widehat{K},d)- L_2(\widehat{K},d)$, where $C_{\max}^*(\widehat{K},d)$ is the length of schedule $S_d$.

After that, we start the loop on $i'$ in which we will adjust the current list ${\cal F}(\widehat{K})$ of solutions for configuration $\widehat{K}$. (Before starting the loop on $d$, the list consists of the single dummy solution $\tilde F=(0,\infty)$.) At each $i'$, for the p-c $\widehat{K}'=(K',i')$, we enumerate its Pareto-optimal solutions $\bar F'=(F_1',F_2')\in{\cal F}(\widehat{K}')$ in the ascending order of $F_1'$ (and the descending order of $\Delta(\bar F')=F_2'-F_1'$). Given a solution $\bar F'$ and schedule $S_d$, we form a solution $\bar F''=(F_1'',F_2'')$ for configuration $\widehat{K}$ as follows.

$F_1'':=F_1' +\rho_{i',i^*}+ L_1(\widehat{K},d)$.

$F_2'':= \left\{\begin{array}{ll}
              F_2'+\rho_{i',i^*}+ L_2(\widehat{K},d), & \mbox{if $\Delta(\bar F')\ge\delta(\widehat{K},d)$}\ (\mbox{\em a solution of type $(a)$});\\
              F_1'+\rho_{i',i^*}+ C_{\max}^*(\widehat{K},d), & \mbox{if $\Delta(\bar F')<\delta(\widehat{K},d)$\ (\mbox{\em a solution of type $(b)$}).}
              \end {array} \right. $

\medskip Case $(b)$ means that the component $F_2'$ does not affect the parameters of the resulting solution $\bar F''$ any more, and so, considering further solutions $\bar F'\in{\cal F}(\widehat{K}')$ (with greater values of $F_1'$ and smaller values of $\Delta(\bar F')$) makes no sense, since it is accompanied by a monotonous increasing of both $F_1''$ and $F_2''$ (between which, a constant difference is established equal to $C_{\max}^*(\widehat{K},d)-L_1(\widehat{K},d)$). Thus, for any given p-c $\widehat{K}'$, a solution of ``type (b)'' can be obtained at most once.

For each solution $\bar F''$ obtained, we immediately try to understand whether it should be \textbf{added} to the current list ${\cal F}(\widehat{K})$, and if so, whether we should remove some solutions from list ${\cal F}(\widehat{K})$ (majorized by the new solution $\bar F''$).

To get answers to these questions, we find a solution $\bar F^{\ell}=(F^{\ell}_1,F^{\ell}_2)$ in list ${\cal F}(\widehat{K})$ with the maximum value of the component $F^{\ell}_1$ such that $F^{\ell}_1\le F_1''$. Such a solution always exists (we call it a \emph{control element} of list ${\cal F}(\widehat{K})$). Since in the loop on $\bar F'\in{\cal F}(\widehat{K}')$, component $F_1''$ monotonously increases, the search for the control element matching $\bar F''$ can be performed not from the beginning of list ${\cal F}(\widehat{K})$, but from the current control element. Before starting the loop on $\bar F'$, we assign the first item of list ${\cal F}(\widehat{K})$ to be the current control element.

If the inequality $F^{\ell}_2\le F_2''$ holds, the current step of the loop on $\bar F'\in{\cal F}(\widehat{K}')$ ends \textbf{without including} the solution $\bar F''$ in list ${\cal F}(\widehat{K})$ (we pass on to the next solution $\bar F'\in{\cal F}(\widehat{K}')$). Otherwise, if $F_2''< F^{\ell}_2$, we look through list ${\cal F}(\widehat{K})$ (starting from the control element $\bar F^{\ell}$) and remove from the list all solutions $\bar F=(F_1,F_2)$ majorized by the new solution $\bar F''$ (which is expressed by the relations $F_1''\le F_1,\ F_2''\le F_2$). At that, the condition $F_1''=F_1$ is sufficient for removing the current control element, while the inequality $F_2''\le F_2$ is sufficient for removing subsequent elements. The scanning of list ${\cal F}(\widehat{K})$ stops as soon as either the first non-majorized list item is found distinct from the control element (for this item and for all subsequent items, the relations $F_2<F_2''$ hold), or if the list has been scanned till the end. Include solution $\bar F''$ in list ${\cal F}(\widehat{K})$ and assign it to be a new \emph{control element}, which completes the current step of the c-loop.

\section{The analysis of algorithm ${\cal A}_{DP}$}\label{se4}

\begin{theorem}\label{th3}
Algorithm ${\cal A}_{DP}$ finds an optimal solution of problem $\RG 2||C_{\max}$ in time $O(n^{g^2+1})$.
\end{theorem}
\proof\ %
Since the optimality of the solution found by algorithm ${\cal A}_{DP}$ follows explicitly from the properties of the optimal solution proved in Theorems \ref{th1} and \ref{th2}, to complete the proof of Theorem \ref{th3}, it remains to show the validity of bounds on the running time of the algorithm; to that end, it is sufficient to estimate the running time ($T_{BS}$) of the \emph{Main stage} of the algorithm.

In the Main stage, for each basic configuration $\widehat{K}$, the set ${\cal F}(\widehat{K})$ of all its Pareto-optimal solutions is found.
Since this set is formed from the solutions obtained in the previous steps of the algorithm for various pre-configurations of configuration $\widehat{K}$, the obvious upper bound on the value of $T_{BS}$ is the \textbf{product} of the number of configurations ($N_C$), of the number of pre-configurations ($N_{PC}$) for a given configuration, and of the bound ($T_{step}$) on the running time of any step of the loop on configurations and pre-configurations (called a \emph{c-loop}).

In each step of the c-loop, list of solutions ${\cal F}(\widehat{K}')$ of a given p-c $\widehat{K}'$ is scanned. From each such solution, a solution for configuration $\widehat{K}$ is generated which is then either included or not included in list ${\cal F}(\widehat{K})$. The solutions included in the list \textbf{in this step of the c-loop} will be called ``new'' ones; other solutions, included in ${\cal F}(\widehat{K})$ \textbf{before starting this step} will be called ``old''.

While estimating a new solution claiming to be included in ${\cal F}(\widehat{K})$, we scan some ``old'' solutions of list ${\cal F}(\widehat{K})$, which is performed in two stages. In the first stage, we look through the elements from ${\cal F}(\widehat{K})$, starting from the current \emph{control element}, in order to find a \textbf{new control element} immediately preceding the applicant. In the second stage (in the case of the \textbf{positive decision} on including the applicant in the list), we check the (new) control element and the subsequent elements from ${\cal F}(\widehat{K})$ subject to their \textbf{removal from the list} (if they are majorized by the applicant). We continue this process until we find either the first undeletable element or the end of the list. We would like to know: \textbf{how many views} of items of list ${\cal F}(\widehat{K})$ will be required in total in one step of the c-loop? It is stated that no more than $O(Z)$, where $Z$ is the maximum possible size of list ${\cal F}(\widehat{K})$ in any step of the algorithm for all possible configurations~$\widehat{K}$.

To prove this statement, we first note that none of the ``new'' elements included in list ${\cal F}(\widehat{K})$ in this step of the c-loop will be deleted in this step, since all ``new'' solutions included in the list are incomparable by the relation $\lessdot$. This follows from the facts that: 1) all applicants formed by type (a) are incomparable; 2) if the last solution is formed by type (b), then it is either incomparable with the previous applicant, or majorized by it (and therefore, is not included in the list). Thus, only ``old'' elements will be deleted from the list, and the total (in the c-loop step) number of such deletions does not exceed $Z$.

In addition, the viewing of an element from ${\cal F}(\widehat{K})$, when it receives the status of a ``control element'', occurs at most once during each c-loop step, and so, the total number of such views in one step does not exceed $Z$. There may be also ``idle views'' of elements subject to assigning them the status of a ``control element''. Such an idle view may happen only once for each applicant, and so, the number of such idle views during one step of the c-loop does not exceed $|{\cal F}(\widehat{K}')|\le Z$. 

Next, the total (over a step of the c-loop) number of views of elements from ${\cal F}(\widehat{K})$ subject to their removal from the list does not exceed $O(Z)$, as well. Indeed, viewing an element of ${\cal F}(\widehat{K})$ \textbf{with its removal} occurs, obviously, for each element at most once (or, in total over the whole step, at most $Z$ times). Possible ``idle view'' of an element from ${\cal F}(\widehat{K})$ (without its deleting) happens at most once for each applicant, which totally amounts (over the current step of the c-loop) at most $|{\cal F}(\widehat{K}')|\le Z$. Thus, the total number of views of items from ${\cal F}(\widehat{K})$, as well as the total running time of the c-loop step ($T_{step}$), does not exceed $O(Z)$. Let us estimate now number~$Z$ itself.

We know that for any given configuration $\widehat{K}$ the solutions $\bar F=(F_1,F_2)$ from list ${\cal F}(\widehat{K})$ are incomparable with respect to relation $\lessdot$. Thus, the number of elements in list ${\cal F}(\widehat{K})$ does not exceed the number of different values of the component $F_1$. The value of the component $F_1$ is the sum of the workload of machine $A$ and the total duration of its movement. (There are no idle times of machine $A$ in the optimal schedule.) Since the workload of machine $A$ (for a fixed configuration $\widehat{K}$) is fixed, the number of different values of the component $F_1$ can be bounded above by the number of different values that the length of a machine route along the nodes of network $G$ can take. As we know, each passage of the machine along the arc $(v_i, v_j)$ is associated with the performance of at least one job located at node $v_j$. Thus, any machine route contains $x\le k_j\le n_j$ arcs entering node $v_j$, and the same number of arcs ($x$) leaving the node.

Let us define a \emph{configuration of a machine route} as a matrix $H=(h_{ij})$ of size $g\times g$, where $h_{ij}\ (i\ne j)$ specifies the multiplicity of passage of an arc $(v_i,v_j)\in G$ in the route; $h_{jj}=n_j-\sum_{i\ne j} h_{ij}$. Thus, for any $j\in[g]$, the equality holds:
\begin{equation}\label{7}
\sum_{j=1}^g h_{ij}=n_i.
\end{equation}
Clearly, for any closed route the following equalities are also valid:
\begin{equation}\label{8}
\sum_{i=1}^g h_{ij}=n_j,\ \ j\in[g].
\end{equation}
Hence, it follows that the number of different values of the route length of a machine does not exceed the number of configurations of a closed route. The latter does not exceed the number of different matrices $H$ with properties (\ref{7}) and (\ref{8}). Let us (roughly) estimate from above the number ($Z'$) of such matrices without taking into account property (\ref{8}).

The number of variants of the $i$th row of matrix $H$ does not exceed the number of partitions of the number $n_i$ into $g$ parts, i.e., is not greater than
\begin{eqnarray*}
C_{n_i+g-1}^{g-1}&=&\frac{1}{(g-1)!}(n_i+1)(n_i+2)\dots(n_i+g-1)\\
&=& \frac{n_i^{g-1}}{(g-1)!}\left(1+\frac{1}{n_i}\right)\left(1+\frac{2}{n_i} \right)\dots \left(1+\frac{g-1}{n_i}\right)\le \frac{n_i^{g-1}}{(g-1)!}\exp{\frac{(g-1)g}{2n_i}}.
\end{eqnarray*}
Since the value of $\sup_{n_i\in[1,\infty)}\exp{\frac{(g-1)g}{2n_i}}$ depends only on $g$, we obtain an upper bound of the form
$$C_{n_i+g-1}^{g-1}\le f(g)n_i^{g-1}.$$
Denote $\Pi=n_1n_2\dots n_g$. Then $Z\le Z'\le(f(g))^g\cdot \Pi^{g-1}$, and the number of configurations ($N_C$) can be bounded above by $O(g\Pi)$. Finally, the number of pre-configurations is bounded by $N_{PC}\le O(gn)$. Taking into account the above bounds, the bound $\Pi\le n^g/g^g$, and the boundedness of the parameter $g$ by a constant, we obtain the final bound on the running time of the algorithm:
$$T_{{\cal A}}\approx T_{BS}\approx N_C N_{PC} T_{step}\le \varphi(g)\cdot O(\Pi^gn)\le O(n^{g^2+1}).$$
Theeorem \ref{th3} is proved.\qed

\section{Conclusion}\label{concl}

We have considered the two-machine routing flow shop problem on an asymmetric network ($\RG 2||C_{\max}$). We have improved the result by Yu {\it et al.} \cite{YuZhWaFa11} by showing that for a more general problem (the problem with an arbitrary asymmetric network) the property of existing an optimal permutation schedule also holds. Next, we have presented a polynomial time algorithm for the problem with a fixed number of nodes, which is the first positive result on the computational complexity of the general $\RG 2||C_{\max}$ problem.  

We now propose a few open questions for future investigation.

\medskip
{\bfseries Question 1.} What is the parametrized complexity of problem $\RG 2||C_{\max}$ with respect to the parameter $g$?

{\bfseries Question 2.} Are there any subcases of problem $\RG 2||C_{\max}$ with unbounded $g$ (e.g., $G$ is a chain, or a cycle, or a tree of diameter 3, or a tree with a constant maximum degree, \emph{etc.}) solvable in polynomial time?

{\bfseries Question 3.} Are there any strongly NP-hard subcases of problem $\RG 2||C_{\max}$ for which NP-hardness is not based on the underlying \emph{TSP}? In other words, is it possible that for some graph structure $G=W$ the \emph{TSP} on $W$ is easy, but problem $\RG 2|G=W|C_{\max}$ is strongly NP-hard?

 \bibliographystyle{splncs04}
 \bibliography{CherKonSev}

\appendix
\section{The proof of Theorems \ref{th1} and \ref{th2}}\label{se2}

{\em Proof of Theorem \ref{th1}.}
For the reasons of convenience, \textbf{in the proof of this theorem only} we will assume that each job is located at a separate job node. This enables us to assume that each node is visited by each machine only once, and thus the route of each machine in this model is a \emph{Hamiltonian path} in a directed network $G^*=(V^*,E^*)$ from node 0 to node $(n+1)$, or a permutation $\pi=(\pi_0,\pi_1,\dots,\pi_{n+1})$ of indices from 0 to $n+1$ starting with $\pi_0=0$ and ending with index $\pi_{n+1}=n+1$. (The set of such permutations will be denoted as ${\cal P}$.) At that, network $G^*$ admits arcs of zero length.

Let $i\prec_{\pi} j$ and $i\preceq_{\pi} j$ denote strict and non-strict precedence of node $i$ to node $j$ in a route $\pi$ (in particular, $i\preceq_{\pi} j$ admits $i=j$). We will assume that job $j\in[n]$ is located at the node with the same index $j$. Given a permutation $\pi\in {\cal P}$, $V(\pi,j)\doteq\{i>0\,|\,i\preceq_{\pi} j\}$ and $V'(\pi,j)\doteq\{i>0\,|\,i\prec_{\pi} j\}$ will denote the sets of jobs located in the initial segments of sequence $\pi$ including or excluding job $j$, respectively. 

In the course of the proof, we will construct \emph{dense schedules} $S_D\langle r;\pi',\pi''\rangle$ determined by three parameters: a time moment $r\ge 0$ and permutations $\pi',\pi''\in{\cal P}$ specifying the routes of machines $A$ and $B$. The set of such schedules will be denoted as ${\cal S}_D$. Each schedule $S_D\langle r;\pi',\pi''\rangle\in {\cal S}_D$ is constructed according to the following rules: machines $A$ and $B$, starting from node 0 at moments 0 and $r$, respectively, follow (\textbf{without any idle times}) the routes specified by the permutations $\pi'$ and $\pi''$, spending all the time just for processing the jobs and moving between nodes.

It is clear that schedule $S_D\langle r;\pi',\pi''\rangle\in {\cal S}_D$ can be infeasible for some values of parameters $r,\pi',\pi''$. At that, for any pair of permutations $\pi',\pi''\in {\cal P}$ there exists such a value $r=\hat r$ for which the corresponding schedule $S_D\langle \hat r;\pi',\pi''\rangle$ is feasible, and its length coincides with the length of the active schedule $S(\pi',\pi'')$. It is also clear that $\hat r$ is the minimum possible value of $r$ for which schedule $S_D\langle r;\pi',\pi''\rangle$ is feasible. Such a value $\hat r$ is uniquely defined for any given pair $(\pi',\pi'')$; this function will be denoted as $\hat r(\pi',\pi'')$.

In fact, schedule $S_D\langle r;\pi',\pi''\rangle$ is feasible, \emph{iff} machine $B$ arrives at each node $i\in[n]$ not earlier than machine $A$ completes its operation of job $i$:
\begin{equation}\label{1}
r+R(\pi'',j)+B(V'(\pi'',j))\ge R(\pi',j)+A(V(\pi',j)),\ j\in[n],
\end{equation}
where $A(Y),B(Y)$ denote the total length of operations of machines $A$ and $B$ over the jobs from set $Y$, $R(\pi,j)=\sum_{i\in[k]}\rho(\pi_{i-1},\pi_{i})$ is the length of path $(\pi_0,\pi_1,\dots,\pi_k)$ from the depot to node $\pi_k=j$.

Given an instance $I$ of problem $\RG 2||C_{\max}$, let $S$ be such an \textbf{optimal schedule} in which machine~$B$ follows \textbf{the shortest route} around network nodes (among all routes of machine $B$ \textbf{in optimal schedules}). Let $\pi^1,\pi^2\in {\cal P}$ be the routes of machines $A$ and $B$ in that schedule~$S$;\ $\hat r\doteq\hat r(\pi^1,\pi^2)$ and $\widehat S\doteq S_D\langle \hat r; \pi^1,\pi^2\rangle$. Then schedule $\widehat S\in {\cal S}_D$ is feasible and optimal.

Let us number the nodes of network $G^*$ (as well as jobs) according to the order of their passing by machine $A$: nodes are numbered by indices from 0 to $n+1$, and jobs by indices from 1 to $n$. Thus, $\pi^1=(\pi^1_0,\pi^1_1,\dots,\pi^1_{n+1})=(0,1,\dots,n+1)$. Let us define in sequence $\pi^2$ a sub-sequence of \emph{marked nodes} $\pi^*=(\pi^2_{\nu_0},\pi^2_{\nu_1},\dots,\pi^2_{\nu_T})$ by the recursion:\ $\nu_0=0$,\ $\nu_t=\min\{j\,|\,\pi^2_j > \pi^2_{\nu_{t-1}}\}, t\in[T];\ \nu_T=n+1$. In other words, we go along the route of machine $B$ and ``mark'' the nodes according to a simple algorithm: first, we mark node~0; next, we mark the first met node with a larger index, and so on, until we arrive at node $(n+1)$ (which we also mark). Then we have: $0=\nu_0<\nu_1(=1)<\dots<\nu_T=n+1$ and $0=\pi^2_{\nu_0}<\pi^2_{\nu_1}<\dots<\pi^2_{\nu_{T-1}}(=n)< \pi^2_{\nu_{T}}=n+1$. It can be also easily seen that $\pi^*$ is a sub-sequence of both sequences: $\pi^1$ and $\pi^2$.

Let $L$ denote the set of marked nodes. Other nodes will be called \emph{mobile} ones. We denote by $W^1_t,W^2_t\ (t\in T)$ the sets of mobile nodes being passed by machines $A$ and $B$ between two consecutive marked nodes: $\pi^2_{\nu_{t-1}}$ and $\pi^2_{\nu_{t}}$. (These sets will be referred to as \emph{segments} of permutations $\pi^1$ and $\pi^2$.) Clearly, $W^2_1=W^1_T=\varnothing$.

While speaking on the difference between permutations $\pi^1$ and $\pi^2$, it can be observed that each mobile node is located in $\pi^1$ in a segment with a lesser index than in $\pi^2$. (For example, all elements of $W^2_2$ come there from $W^1_1$.) Based on this property of permutations $\pi^1,\pi^2$, procedure \emph{Trans} described below transforms the route of machine $B$ step by step, transferring exactly one mobile node to a new position in each step. In the course of this transformation, the current (variable) permutation specifying the route of machine $B$ will be denoted by $\tilde\pi$. Since this transformation of the route of machine $B$ leaves the mutual order of the marked nodes stable, we can transfer the above definition of segments (of permutations $\pi^1$ and $\pi^2$) onto permutation $\tilde\pi$.

At the end of procedure \emph{Trans}, the route of machine $B$ will coincide with the route of machine $A$ (i.e., we will have $\tilde\pi=\pi^1$), and the corresponding dense schedule $S(\tilde\pi)\doteq S_D\langle\hat r;\pi^1,\tilde\pi\rangle$ will become a permutation one. After completing the description of procedure \emph{Trans}, Lemma \ref{le1} is proved providing some important properties of schedules $S(\tilde\pi)$ obtained in steps of the procedure.

\begin{center}
\bf Procedure \trans
\end{center}

Procedure is divided into $T$ \emph{stages}, where in the $t$th stage ($t\in[T]$) we consider the transfer of all mobile nodes from the $t$th segment of permutation $\pi^2$ to their ``proper places'', i.e., to those segments where they stand in permutation $\pi^1$, and in the \textbf{ascending order of their numbers}. (The first stage is, therefore, empty, since $W^2_1=\varnothing$.) The $t$th stage is divided into \emph{steps}, where in each step the transfer of the \emph{current mobile node} standing in the current permutation $\tilde\pi$ at position $\nu_t-1$ is performed. Clearly, each such transference of a node to one of the preceding segments reduces by 1 the number of mobile nodes in the \emph{current} ($t$th) segment of permutation $\tilde\pi$, and so, after a finite number of steps, we will see in this position the \textbf{marked node} $\pi^2_{\nu_{t-1}}$, which means the completion of the stage.\algend

\medskip We notice that in any step of stage $t$, all nodes preceding in $\tilde\pi$ the marked node $\pi^2_{\nu_{t-1}}$ (inclusively) are sequenced in $\tilde\pi$ in the \textbf{ascending order of their numbers}.

\begin{lemma}\label{le1}
A dense schedule $S(\tilde\pi)$ obtained after each step of procedure \emph{\trans} is feasible. At that, machine $B$ arrives at each marked node in schedule $S(\tilde\pi)$ not earlier (and by a not shorter way) than in schedule $\widehat S$.
\end{lemma}
\proof\ %
Since we start in \trans\ from permutation $\tilde\pi:=\pi^2$, the corresponding schedule $S(\tilde\pi)=\widehat S$ is feasible. Suppose that in some step of stage~$t$, schedule $S(\tilde\pi)$ is still feasible, and let an item $\tilde\pi_j\doteq y$ of $\tilde\pi$ be transferred from position $j=\nu_t-1$ to position $i<j$. (At the same time, all items $\{\tilde\pi_i,\dots, \tilde\pi_{j-1}\}$ increase their positions in $\tilde\pi$ by 1.) For convenience, we leave the notation $\tilde\pi$ for the route of machine $B$ \textbf{before} this transposition, while the new permutation of nodes (\textbf{after} the transposition of node $y$) will be denoted by~$\tilde\pi'$.

Note that $\tilde\pi'_{i-1} (=\tilde\pi_{i-1})< y < \tilde\pi'_{i+1} (=\tilde\pi_i)$. Suppose that in schedule $S(\tilde\pi)$ machine $B$ leaves node $\tilde\pi_{i-1}$ at time $\tau$. Then it appears at node $\tilde\pi_i$ at time
\begin{equation}\label{2}
\tau+\rho(\tilde\pi_{i-1},\tilde\pi_i)\ge \tau'
\end{equation}
(where $\tau'$ is the time when machine $A$ leaves node $\tilde\pi_i$), since schedule $S(\tilde\pi)$ is feasible. What will change in the schedule after the transposition of node $y$ to the position between nodes $\tilde\pi_{i-1}$ and $\tilde\pi_i$? In the new permutation $\tilde\pi'$, machine $B$ goes from node $\tilde\pi_{i-1}$ (at time $\tau$) to node $y$, and only after that to node $\tilde\pi_i$. Suppose that schedule $S(\tilde\pi')$ is infeasible with respect to job $y$. This means that in this schedule machine $B$ arrives at node $y$ too early --- when machine $A$ has not yet completed this job. Suppose, it needs $\varepsilon>0$ time to complete the job. Then machine $A$ will exit from node $y$ at time $\tau+\rho(\tilde\pi_{i-1},y) +\varepsilon$ and will arrive at node $\tilde\pi_i$ not earlier than by time $\tau+\varepsilon+\rho(\tilde\pi_{i-1},y) +\rho(y,\tilde\pi_{i})$. Hence, $\tau'=\tau+\varepsilon+\rho(\tilde\pi_{i-1},y)+\rho(y,\tilde\pi_{i})+\Delta$, where $\Delta$ is a non-negative additive (including the length of the operation of job $\tilde\pi_i$ on machine $A$). In view of the triangle inequality, we obtain the inequality $\tau'>\tau+\rho(\tilde\pi_{i-1},\tilde\pi_i)$, contradicting~(\ref{2}). This implies that schedule $S(\tilde\pi')$ is \textbf{feasible for job} $y=\tilde\pi'_i$. It remains feasible for jobs $\tilde\pi'_{i+1},\dots,\tilde\pi'_j$ as well, since the insertion of job $y$ prior to these jobs in the route of machine $B$ just postpones their execution by machine $B$ to a later time (compared to that in schedule $S(\tilde\pi)$).

In the subsequent positions ($j+1=\nu_t$ and onwards), sequence $\tilde\pi$ has not changed (and coincides with $\pi^2$). Let us show that for jobs in these positions schedule $S(\tilde\pi')$ is also feasible. To that end, it is sufficient to show that in schedule $S(\tilde\pi')$, machine $B$ arrives at node $z\doteq\tilde\pi'_{\nu_t}=\pi^2_{\nu_t}$ (and therefore, at the subsequent nodes, as well) not earlier than in $\widehat S$. Indeed, if to assume that it arrives at node $z$ in schedule $S(\tilde\pi')$ earlier than in $\widehat S$ (at that, having processed the same set of jobs $V'(\tilde\pi',z)= V'(\pi^2,z)$), this would mean that its path from the depot to node $z$ is \textbf{shorter} in the route $\tilde\pi'$ than in $\pi^2$. In this case, we could define a new schedule $\widehat S'$ being a combination of schedules $S(\tilde\pi')$ and $\widehat S$. For machine $A$, it remains the same as in both schedules; for machine $B$, we take schedule $S(\tilde\pi')$ for jobs from $V'(\tilde\pi',z)=V'(\pi^2,z)$ and schedule $\widehat S$ for the remaining jobs. Clearly, schedule $\widehat S'$ is not dense. However, it would be feasible and \textbf{optimal}, since its length coincides with that of schedule $\widehat S$. In addition, as we established, the route $\tilde\pi'$ of machine $B$ is shorter than in schedule $\widehat S$, which contradicts the choice of schedule $\widehat S$. Thus, the feasibility of schedule $S(\tilde\pi')$ obtained after performing the current step of procedure \trans\ is confirmed. At the same time, we have proved that machine $B$ arrives in schedule $S(\tilde\pi')$ at the marked node $z$ (and hence, at the subsequent marked nodes) not earlier than in schedule $\widehat S$. It is clear that during the subsequent steps of the \trans\ procedure this property will not be violated, since during the subsequent transpositions only \textbf{additional nodes will be inserted} prior to node~$z$. Lemma~\ref{le1} is proved.\qed
\medskip\indent %
Let us proceed with the proof of Theorem \ref{th1}. Applying Lemma \ref{le1} to permutation $\tilde\pi=\pi^1$ obtained \textbf{at the completion} of procedure \trans, we obtain:
$$R(\pi^2,j)\le R(\pi^1,j),\ \forall\ j\in L.$$
Applying these inequalities to relations (\ref{1}) for $j\in L$, we derive the inequalities:
\begin{equation}\label{3}
\hat r+B(V'(\pi^2,j))\ge A(V(\pi^1,j)),\ j\in L.
\end{equation}
Now, let us consider an arbitrary $j\in [n]$, and let $z\doteq\pi^2_{\nu_t}\preceq_{\pi^2} j\prec_{\pi^2}\pi^2_{\nu_{t+1}}$. Then
\begin{equation}\label{4}
\hat r+B(V'(\pi^2,j))\ge \hat r+B(V'(\pi^2,z)) \twofl{(\ref{3})}{\ge} A(V(\pi^1,z))\ge A(V(\pi^2,j)).
\end{equation}
The last inequality in this chain follows from the fact that all nodes standing in permutation $\pi^2$ prior to $j$ (inclusively) have indices not greater than $z$. (It is evident for marked nodes; the mobile nodes, appeared in permutation $\pi^2$ in the $(t+1)$th or preceding segments, precede to the marked node $z$ in permutation $\pi^1$, whence $V(\pi^2,j)\subseteq V(\pi^1,z)$.)

And now the final! Consider the permutation schedule $S^*\doteq S_D\langle\hat r;\pi^2,\pi^2\rangle$ in which machine $A$ follows the route $\pi^2$. Then both machines pass the same ways to any node $j$. Adding the length of that path to both parts of the resulting inequality (\ref{4}), we obtain relations:
$$\hat r+R(\pi^2,j)+B(V'(\pi^2,j))\ge R(\pi^2,j)+ A(V(\pi^2,j)),\ j\in[n],$$
sufficient for the feasibility of schedule $S^*$ (see (\ref{1})). Besides, since the schedule of machine $B$ in $S^*$ coincides with that in schedule $\widehat S$, both schedules have the same length, which means that schedule $S^*$ is also optimal. Theorem \ref{th1} is proved.\qed
\medskip\indent Now we return to the original model on network $G$, where we will consider \textbf{permutation schedules} only. A permutation $\pi=(\pi_1,\dots,\pi_n)$ of job indices of a given instance $I$ will specify the order of job processing by both machines, which also uniquely defines the routes of the machines. Function $\tilde S(I,\pi)$ will define an active schedule, specified for a given instance~$I$ by permutation $\pi$.

\begin{definition}\label{de2}\rm
Farther, for the convenience of arguing, we renumber the jobs $J_j\ (j\in[n])$ in the descending order of their Johnson's priorities (see Definition~\ref{de1}). Given a permutation $\pi=(\pi_1,\dots,\pi_n)$ of job indices and a node $v\in V$, we will say that a pair of jobs $\pi_i,\pi_j\in{\cal J}(v),\ i<j,$ \emph{stays improperly} (and forms an \emph{inversion}) in $\pi$, if $\pi_i> \pi_j$; \emph{Inv}$(\pi)$ will denote the total (over all $v\in V$) number of such pairs in $\pi$. Jobs \mbox{$\pi_i,\pi_j\in{\cal J}(v),\ i<j$}, will be called \emph{$v$-neighbors}, if there are no other jobs from ${\cal J}(v)$ between them in $\pi$.
\end{definition}

{\em Proof of Theorem \ref{th2}.}
Among the set of permutations defining optimal permutation schedules (this set is nonempty, by Theorem \ref{th1}), we choose the permutation $\pi^*$ on which the minimum of the function \emph{Inv}$(\pi)$ is attained, and show that \emph{Inv}$(\pi^*)=0$.

Suppose that \emph{Inv}$(\pi^*)> 0$. Then in network $G$ there is a node $v^*\in V$ and a couple of jobs in this node standing in permutation $\pi^*$ \emph{improperly}, and so, there are also $v^*$-neighbors in $\pi^*$ with this property: $\pi^*_i>\pi^*_j,\ i<j$. To show that this is impossible, we transform the original instance $I$ of problem $\RG 2$ into an instance $\hat I$ of problem $F2$ as follows.

Let $S^*=\tilde S(I,\pi^*)$. In this active schedule, we will distinguish the time intervals of processing the jobs from ${\cal J}(v^*)$. All other time intervals of ``not working at node $v^*$'' (where a machine either performs jobs not from ${\cal J}(v^*)$, or moves between network nodes, or simply stands idle) will be called ``inserts''. (In fact, ``inserts'' will stand for such maximal by inclusion time intervals for each machine.) Due to the permutability of schedule $S^*$, there is a one-to-one correspondence between the inserts of machines $A$ and $B$. They are divided into pairs related to processing the same subsets of jobs not lying at~$v^*$. We represent each such pair of inserts in the form of a schedule of some \emph{pseudo-job}. (These pseudo-jobs will be assigned to additional indices $k\in{\cal K}$.) For such a schedule to be feasible, it is necessary for each pseudo-job to get rid of the overlapping of its ``operations''. 

To that end, we do the following \emph{transformation} of schedule $S^*$: if for some pseudo-job $k'\in{\cal K}$, the intervals $[s_{1k'},s_{1k'}+a_{k'}]$ and $[s_{2k'},s_{2k'}+b_{k'}]$ of processing its ``operations'' on machines $A$ and $B$ have an intersection of length $\lambda_{k'}>0$, then we reduce the lengths of these operations (and synchronously, the moments of their completion) by the amount~$\lambda_{k'}$; new operation durations are: \ $a_{k'}':=a_{k'}-\lambda_{k'};\ b_{k'}':=b_{k'}-\lambda_{k'}$. And farther, instead of the pseudo-job $k'$, we will consider a ``new job'' with the same index $k'$ and the ``shortened'' lengths of the operations: $(a_{k'}',b_{k'}')$. The intervals of processing these operations do not overlap in the modified schedule, and so, the schedule becomes feasible with respect to job~$k'$. It is also clear that such a transformation does not violate the feasibility of the schedule for other jobs (since the schedule for all jobs following $k'$ is shifted synchronously on machines $A$ and $B$ by the same amount~$\lambda_{k'}$) and does not change the values of $\lambda_k$ already set for other pseudo-jobs. At that, the schedule length is reduced by $\lambda_{k'}$.

We will denote by $\hat I$ the instance of problem $F2$ consisting of the ``new jobs'' and jobs from ${\cal J}(v^*)$; $\widehat S$ will stand for the schedule for this set of jobs, obtained from $S^*$ after performing the reduction of pseudo-jobs. Clearly, it is a  permutation schedule. Let $\hat\pi$ denote the permutation of jobs of instance $\hat I$ in schedule~$\widehat S$.

What else can be said about schedule $\widehat S$? First, that 
\begin{equation}\label{5}
C_{\max}(\widehat S)=C(S^*)-\Lambda,
\end{equation}
where $\Lambda\doteq\sum_{k\in{\cal K}}\lambda_k$. Second, that the schedule is ``absolutely dense'', i.e., contains nothing but the intervals of processing the operations in the whole interval from 0 to $C_{\max}(\widehat S)$. (All machine movements and idle times were ``packed'' into the operations of pseudo-jobs.) Third, it is feasible for jobs of instance $\hat I$ and is optimal (its length coincides with the workload of each machine). Evidently, $\widehat S$ is the active schedule defined by permutation $\hat\pi$, i.e., $\widehat S=\tilde S(\hat I,\hat\pi)$.

Let us transform permutation $\hat\pi$ into a permutation $\hat\pi'$ as follows. It can be seen that the $v^*$-neighbors in permutation $\pi^*$, jobs $i^*=\pi^*_i$ and $j^*=\pi^*_j$ standing there \emph{improperly}, are either adjacent in permutation $\hat\pi$ or separated from each other by a new job $k\in{\cal K}$. In the first case, just swap them. In the second case, it is clear that job $k$ stands improperly with at least one of its neighbors: either $k\prec i^*$ or $j^*\prec k$ (since otherwise, the relation $i^*\prec j^*$ would follow from the transitivity of the relation $\prec$, which would contradict the choice of these jobs). Suppose, for example, that $k\prec i^*$. Then, swap $i^*$ first with $k$ and then with $j^*$. As was shown by Johnson \cite{J53}, rearranging neighboring jobs standing improperly does not increase the length of the schedule. Thus, in both cases, the length of the active schedule $\widehat S'=\tilde S(\hat I,\hat\pi')$ will not exceed the length of schedule~$\widehat S$:
\begin{equation}\label{6}
C_{\max}(\widehat S')\le C_{\max}(\widehat S).
\end{equation}

Finally, we transform schedule $\widehat S'$ into a schedule $S'$ of the original instance $I$ by increasing the durations of the operations of each job $k\in{\cal K}$ by $\lambda_k$ (an ``anti-reduction'' procedure) and by restoring (in the time intervals obtained) the schedule for those jobs of the original instance that were performed in schedule $S^*$ within these pseudo-jobs. What can we say about schedule~$S'$?

Firstly, it is \textbf{feasible}. For jobs from ${\cal J}(v^*)$, this follows from the feasibility of schedule $\widehat S'$ and from the obvious fact that the ``anti-reduction'' procedure of each ``new job'' $k\in{\cal K}$ synchronously shifts the intervals of processing the operations of any subsequent job $j\in {\cal J}(v^*)$ on machines $A,B$ by the same amount, which preserves the feasibility of the schedule for job $j$. Let us show the feasibility for the remaining jobs of the original instance $I$ (not belonging to ${\cal J}(v^*)$, and thus, included in some pseudo-jobs). 

We will show this for jobs included in a pseudo-job $k'\in{\cal K}$. If the ``operations'' of this pseudo-job do not overlap in schedule $S^*$ (and thus, they have not undergone the reduction procedure while the transformation $S^*\rightarrow\widehat S$ of the original schedule), then they do not overlap in schedule $\widehat S'$, as well, since the latter is feasible. Clearly, the anti-reduction procedure (implemented to other pseudo-jobs) has no impact on the mutual positions of the two operations of job $k'$ (they do not overlap in schedule $S'$, as before), which implies the feasibility of $S'$ with respect to pseudo-job $k'$, as well as for all original jobs included in $k'$.

Let us next assume that the ``operations'' of pseudo-job $k'$ have undergone the reduction procedure. After the reduction of these ``operations'', a minimum possible gap is established between the starting times of the reduced operations of the new job $k'$ (it is minimum possible, since it coincides with the length of the $A$-operation of this job). In the course of further transformations of the schedule, consisting in a series of transpositions of neighboring jobs and resulted in schedule $\widehat S'$, this gap could not decrease, since schedule $\widehat S'$ is feasible. Finally, the last transformation of the schedule (from $\widehat S'$ to $S'$) preserves all the gaps achieved in the previous stage, because the ``anti-reduction'' procedure can only shift the starting times of both operations of each job by the same amount. Thus, we may conclude that the gap between the starting times of two operations of pseudo-job $k'$ in schedule $S'$ cannot be less than in schedule $S^*$. Now, for proving the feasibility of schedule $S'$ with respect to all jobs included in pseudo-job $k'$, it is sufficient to take into account the following simple facts:

1) The original schedule $S^*$ was feasible with respect to these jobs.

2) The ``operations'' of pseudo-job $k'$ have the same durations in schedules $S^*$ and $S'$.

3) Relative positions of the operations of real jobs within the corresponding ``inserts'' remain the same in $S'$ as in $S^*$.

So, we have proved that schedule $S'$ is feasible.

Next, we have $C(S')=C_{\max}(\widehat S')+\Lambda$, which, together with (\ref{5}) and (\ref{6}), implies the inequality $C(S')\le C(S^*)$ (and therefore, schedule $S'$ is optimal).

And finally, it is clear that the transpositions of neighboring jobs implemented in the course of the transformation of schedule $\widehat S$ to schedule $\widehat S'$ reduce by 1 the number of inversions in the permutation of the original jobs (i.e., for the job permutation $\pi'$ corresponding to schedule $S'$, relation \mbox{\emph{Inv}$(\pi')=$\emph{Inv}$(\pi^*)-1$} holds), which contradicts the choice of permutation~$\pi^*$. The contradiction completes the proof of Theorem \ref{th2}.\qed
\end{document}